\documentclass[twoside,twocolumn,9pt]{article}
\usepackage{extsizes}
\usepackage[super,sort&compress,comma]{natbib} 
\usepackage[version=3]{mhchem}
\usepackage[left=1.5cm, right=1.5cm, top=1.785cm, bottom=2.0cm]{geometry}
\usepackage{balance}
\usepackage{times,mathptmx}
\usepackage{amssymb}
\usepackage{upgreek}
\usepackage{sectsty}
\usepackage{graphicx} 
\usepackage{lastpage}
\usepackage[format=plain,justification=justified,singlelinecheck=false,font={stretch=1.125,small,sf},labelfont=bf,labelsep=space]{caption}
\usepackage{float}
\usepackage{fancyhdr}
\usepackage{fnpos}
\usepackage[english]{babel}
\usepackage{array}
\usepackage{droidsans}
\usepackage{charter}
\usepackage[T1]{fontenc}
\usepackage[usenames,dvipsnames]{xcolor}
\usepackage{setspace}
\usepackage[compact]{titlesec}

\definecolor{cream}{RGB}{222,217,201}

\begin{document}

\pagestyle{fancy}
\thispagestyle{plain}
\fancypagestyle{plain}{

\renewcommand{\headrulewidth}{0pt}
}

\makeFNbottom
\makeatletter
\renewcommand\LARGE{\@setfontsize\LARGE{15pt}{17}}
\renewcommand\Large{\@setfontsize\Large{12pt}{14}}
\renewcommand\large{\@setfontsize\large{10pt}{12}}
\renewcommand\footnotesize{\@setfontsize\footnotesize{7pt}{10}}
\makeatother

\renewcommand{\thefootnote}{\fnsymbol{footnote}}
\renewcommand\footnoterule{\vspace*{1pt}%
\color{cream}\hrule width 3.5in height 0.4pt \color{black}\vspace*{5pt}} 
\setcounter{secnumdepth}{5}

\makeatletter 
\renewcommand\@biblabel[1]{#1}            
\renewcommand\@makefntext[1]%
{\noindent\makebox[0pt][r]{\@thefnmark\,}#1}
\makeatother 
\renewcommand{\figurename}{\small{Fig.}~}
\sectionfont{\sffamily\Large}
\subsectionfont{\normalsize}
\subsubsectionfont{\bf}
\setstretch{1.125} 
\setlength{\skip\footins}{0.8cm}
\setlength{\footnotesep}{0.25cm}
\setlength{\jot}{10pt}
\titlespacing*{\section}{0pt}{4pt}{4pt}
\titlespacing*{\subsection}{0pt}{15pt}{1pt}

\fancyfoot{}
\fancyfoot[RO]{\footnotesize{\sffamily{1--\pageref{LastPage} ~\textbar  \hspace{2pt}\thepage}}}
\fancyfoot[LE]{\footnotesize{\sffamily{\thepage~\textbar\hspace{3.45cm} 1--\pageref{LastPage}}}}
\fancyhead{}
\renewcommand{\headrulewidth}{0pt} 
\renewcommand{\footrulewidth}{0pt}
\setlength{\arrayrulewidth}{1pt}
\setlength{\columnsep}{6.5mm}
\setlength\bibsep{1pt}

\makeatletter 
\newlength{\figrulesep} 
\setlength{\figrulesep}{0.5\textfloatsep} 

\newcommand{\topfigrule}{\vspace*{-1pt}%
\noindent{\color{cream}\rule[-\figrulesep]{\columnwidth}{1.5pt}} }

\newcommand{\botfigrule}{\vspace*{-2pt}%
\noindent{\color{cream}\rule[\figrulesep]{\columnwidth}{1.5pt}} }

\newcommand{\dblfigrule}{\vspace*{-1pt}%
\noindent{\color{cream}\rule[-\figrulesep]{\textwidth}{1.5pt}} }

\makeatother

\twocolumn[
  \begin{@twocolumnfalse}
\vspace{3cm}
\sffamily
\begin{tabular}{m{4.5cm} p{13.5cm} }

& 
\noindent\LARGE{\textbf{Peeling an elastic film from a soft viscoelastic adhesive: experiments and scaling laws.}} \\
\vspace{0.3cm} & \vspace{0.3cm} \\

& \noindent\large{Hugo Perrin$^{\ast}$\textit{$^{a}$}, Antonin Eddi\textit{$^{b\ddag}$}, Stefan Karpitschka$^{c}$,
Jacco H. Snoeijer$^{d}$ and Bruno Andreotti\textit{$^{a\ddag}$}}\\

& 
\noindent\normalsize{The functionality of adhesives relies on their response under the application of a load. Yet, it has remained a challenge to quantitatively relate the macroscopic dynamics of peeling to the dissipative processes inside the adhesive layer. Here we investigate the peeling of a reversible adhesive made of a polymer gel, measuring the relationship between the peeling force, the peeling velocity, and the geometry of the interface at small-scale. Experiments are compared to a theory based on the linear viscoelastic response of the adhesive, augmented with an elastocapillary regularization approach. This theory, fully quantitative in the limit of small surface deformations, demonstrates the emergence of a ``wetting" angle at the contact line and exhibits scaling laws for peeling which are in good agreement with the experimental results. Our findings provide a new strategy for design of reversible adhesives, by quantitatively combining wetting, geometry and dissipation.} \\

\end{tabular}

 \end{@twocolumnfalse} \vspace{0.6cm}

]

\renewcommand*\rmdefault{bch}\normalfont\upshape
\rmfamily
\section*{}
\vspace{-1cm}


\footnotetext{\textit{$^{a}$~Laboratoire de Physique Statistique (LPS), UMR 8550 CNRS, ENS, Univ. Paris Diderot, Sorbonne Universit\'e, 24 rue Lhomond, 75005, Paris, France.}}
\footnotetext{\textit{$^{b}$~Laboratoire de Physique et M\'ecanique des Milieux H\'et\'erog\`enes (PMMH), UMR 7636 CNRS, ESPCI, Univ. Paris Diderot, Sorbonne Universit\'e, 10 rue Vauquelin, 75005 Paris, France}}
\footnotetext{\textit{$^{c}$~Max Planck Institute for Dynamics and Self-Organization, Am Fassberg 17, 37077 Goettingen, Germany}}
\footnotetext{\textit{$^{d}$~Physics of Fluids Group, Faculty of Science and Technology, University of Twente, P.O. Box 217, 7500 AE Enschede, The Netherlands}}

\section{Introduction}
Pressure sensitive adhesives, ubiquitous for domestic and industrial applications, have the characteristic property that they do not undergo chemical reactions during the bonding process and their performance life. Animals with adhesive pads are ubiquitous in nature~\cite{LW14} and have inspired numerous designs of artificial reversibly adhesive materials~\cite{CC16,BGAC10,JH11,SEK16}. The insights combining viscoelasticity, capillarity, and multiscale hierarchical topography~\cite{ALHZCKFF00} are crucial to design innovative adhesives. Namely, an effective adhesive material should stick under physical contact with a substrate and must therefore respond highly compliant, similar to a liquid. Its adhesive performance results from the resistance to peeling it off a substrate. Strong adhesion implies a high energy dissipation, produced in most polymeric materials by fingering instabilities \cite{vilmin2009simple,ghatak2003adhesion,ghatak2000meniscus,nase2008}, cavitation  \cite{teisseire2007understanding} and fibrillar deformation.

From a theoretical perspective, pioneering models have considered the opposite limit of weak adhesion, for which the debonding is interfacial, reversible, and the adhesives remain weakly deformed~\cite{kaelble1960}. According to these theories~\cite{deGen89adh,Gennes1996aa,deGe88,Schapery1975a,Schapery1975b,Hui1992a,HAIAT200369,PREPerson,BF09}, the dissipation during debonding can be related to the linear viscoelastic properties of the adhesive~\cite{Schapery1975b,SOAR04}. The recent review by \citet{CC16} gives a comprehensive overview of the development of the field. Most experiments with peeling adhesive tapes~\cite{barquins1995cinetique,Gent96} or bulk fracture~\cite{Cristiano11} disagree quantitatively with theoretical predictions~\cite{schapery1975theory,rahulkumar2000cohesive,BF09}. This leaves a gap in first principles understanding, and has led to the conclusion that non-linear viscoelastic dissipation and, most often, damage mechanisms in the polymer network should be taken into account~\cite{VC15,CC16}.

\begin{figure}[t!]
\includegraphics{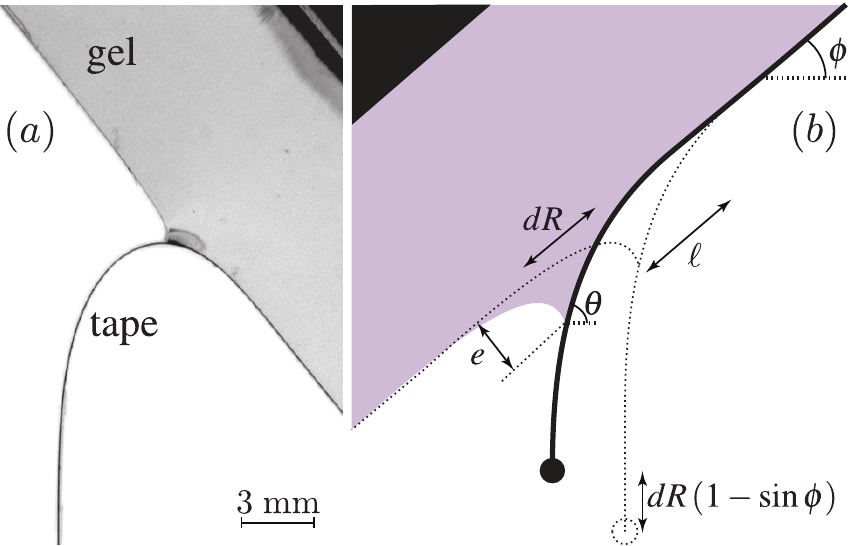}
\caption{(a) Image of the peeling experiment. A tape is pulled vertically from the adhesive gel under the influence of a weight  (not in the image) hanging at its end. (b) Schematic illustrating the gravitational energy released by a peeling a length $dR$. The forcing is controlled by varying the attached mass and the inclination $\phi$. The image also defines the normal deflection $e$, the lateral scale $\ell=(B/G)^{1/3}$ on the side where the tape is still attached, and  the tape inclination at the contact line $\theta$.}
\label{Fig1}
\end{figure}

In this paper we investigate the dynamics of peeling for a reversible viscoelastic adhesive, which can be peeled off without exhibiting irreversible plastic damage (Fig.~\ref{Fig1}). Rather than considering the classical case of peeling a thin, strong adhesive with a flexible backing off a rigid solid~\cite{kaelble1965peel,barquins1995cinetique,Gent96,vilmin2009simple}, we take the opposite perspective: we use a thick layer of a weak adhesive on a rigid backing and peel off a thin flexible tape of a much stiffer material (cf. Fig.~\ref{Fig1}). This way, we disentangle the effects of bending elasticity, viscoelastic dissipation, and adhesion energy. Our model adhesive is made of a multi-scale polymer gel, whose strong dissipation is controlled by linear viscoelasticity.
The key finding is that the dissipation in the bulk is determined by the singular deformation in the vicinity of the contact line, this singularity being regularised by surface energy. This is in stark contrast to ``classical'' peeling where the blunt and frequently irregular crack front cannot localize dissipation sufficiently and the thickness of the adhesive matters~\cite{CC16}. To resolve all our experimental findings in detail, we propose a theory based on which we establish new scaling laws for peeling of such reversible adhesives, with multiplicative factors quantitatively determined in the limit of small deformations.
%

\section{Set-up}
Elastomers are reticulated polymers obtained by cross-linking long polymer chains. The longer the chains between cross-links (or entanglement points), the larger their relaxation timescale and their effective viscosity.  Here we use an addition-cure adhesive that consists of a Silicone gel (Dow Corning CY52-276) whose reticulated polymer network is formed by polymerizing small multifunctional prepolymers. At the gel point, a fractal polymer network forms, which is composed of branches of all lengths between the prepolymer and the size of the sample. Once stirred and degassed, the mixture of the two prepolymers is poured into petri dishes to make gel layers of varying thickness (from $2-10$~mm). The stoichiometric ratio of 1:1 (A:B) provides extra cross-links with respect to the gel point and leads to a finite shear elastic modulus $G \simeq 1.2$~kPa. The linear rheology is very accurately fitted by the simple relation
\begin{equation}\label{eq:gelrheology}
\mu(\omega)=G'(\omega) +iG''(\omega) = G \left[1+ (i\tau\omega)^n \right], 
\end{equation}
with an exponent $n\simeq 0.55$. This bulk rheology obeys Kramers-Kronig relation: both the storage and the loss moduli originate from the same relaxation function $G \left[ 1 +\Gamma(1-n)^{-1}\left( \tau/t\right)^{n} \right]$, where $\Gamma$ is the gamma function. The power law dependence of the loss modulus $G'' \sim \omega^{n}$ reflects the architecture of the polymer network, with a continuum distribution of relaxation times reminiscent of that evidenced at the gelation point -- above a frequency $\sim \tau^{-1}$, the rheology remains the same as that observed at the gelation point \cite{Winter:1986aa}. 

The cross-over timescale, measured $\tau=0.13$~s for our system, is determined by the length of largest branches of polymers in the network. Rheological measurements have shown that the elastic domain of the gel extends at least to a strain of $300\%$ at low frequency and the linear range of the rheology extends up to $100\%-200\%$ with a slight dependence on frequency. Moreover, the rheological response for normal displacements at the surface of the gel is found to present a negligible dependance on an externally applied tangential stretching of the gel sample \cite{SRA2018}, confirming the linear behavior of the gel.

In order to directly test the hypothesis of a dissipation governed by linear viscoelasticity, a ten times stiffer gel has also been used. It was prepared by adding $5\%$ of Sylgard 184 to the aforementioned gel, leading to $G\simeq10\;\rm{kPa}$, $n\simeq0.44$, and $\tau\simeq 8.6\;\rm{ms}$. A thin, virtually inextensible film of bending modulus $B$ is placed on the gel, and is subsequently peeled off at controlled forcing. The experiment is performed by placing the system upside-down, inclined at a variable angle $\phi$ with respect to the horizontal~\cite{K72}, and attaching a mass at the end of the tape (Fig.~\ref{Fig1}). The forcing is varied by four orders of magnitude by using different masses and also by peeling due to the weight of the plastic sheet alone. The sheet's bending modulus $B$ was varied by one order of magnitude (respectively $B\simeq 9.7 \times 10^{-5}\;\rm{J}$ and $\simeq 6.9 \times 10^{-6}\;\rm{J}$) by using two types of tape: a $88\,{\rm \mu m}$ thick biaxially oriented polypropylene  film (BOPP manufactured by Innovia) and a $34\,{\rm \mu m}$ thick Mylar film (Polyethylene terephthalate) coated with Aluminium (PET, manufactured by Toray). These correspond to Young's moduli of the order of $1\;\rm{GPa}$, which is well separated from those of the gels. The tapes are smooth at optical scales --~submicrometer roughness could not be measured. Despite no further annealing was applied, the samples did not present any plastic damage nor any spontaneous curvature. The position of the contact line, where the sheet joins the gel, and the geometry of the ridge formed below this contact line are recorded using a  video camera ($1024\times1024$ pixels with a resolution of $20\,{\rm \mu m/pix}$). 

The key control parameter of the experiment is the peeling force $f$ per unit width, defined as the energy released by gravity when the contact line moves by a unit distance. To quantify this force, we focus on the case where the length $R$ of the freely hanging tape is sufficiently large to be quasi vertical close to its free end. Then, peeling the tape by a length $dR$, the end of the tape moves downward by $dR(1-\sin\phi)$ [c.f. Fig.~\ref{Fig1}(b)]. Hence, we obtain the peeling force per unit width:
\begin{equation}\label{eq:f}
f=  \lambda g (1-\sin \phi),
\end{equation}
where $\lambda$ is the mass at the end of the tape per unit width. In this expression we neglected the weight of the sheet, which can been included leading to a general expression (see~Appendix~\ref{app:weight}).

\section{Geometric characteristics of the peeling front} 
\begin{figure}[t!]
\begin{center}
\includegraphics{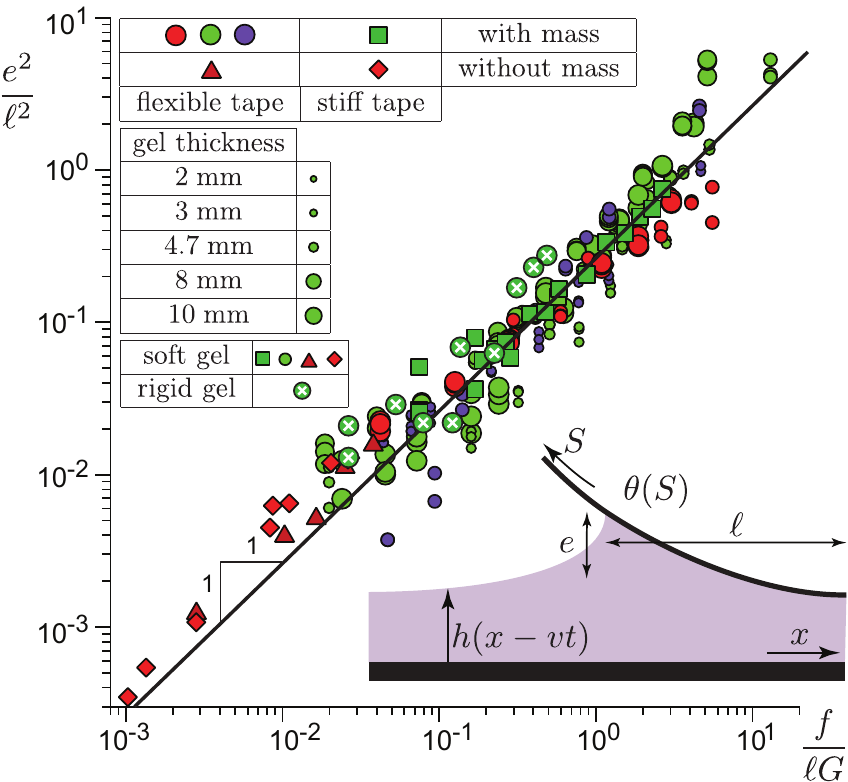}
\caption{Relation between the gel deformation $e$, rescaled by the elasto-bending length $\ell=(B/G)^{1/3}$, and the peeling force $f$ rescaled by $\ell$ and by the gel elastic modulus $G$. The colors correspond to different inclination angles $\phi$ from red to blue ($-45^\circ$, $0^\circ$, $45^\circ$) and the symbols to the type of experiment, as shown in Fig.~\ref{Fig6}. The marker's size codes for gel thickness. The data collapses according to (\ref{eq:normal}) shown as the solid line (prefactor $\simeq 0.26$). The circled cross symbol corresponds to a series of data obtained with a $7.5\;\rm{mm}$ thick layer of the most rigid gel, and with the flexible backing  in the horizontal case $\phi\simeq 0^\circ$.  Inset: the shape of the sheet is parametrised by $\theta$, the shape of the gel by $h$.}
\label{Fig2}
\end{center}
\end{figure}

The present setup provides an experimental access to the geometric features in the vicinity of the contact line \cite{benyahia1997mechanisms}. This is of key importance to understand the viscoelastic dissipation during the peeling, which occurs as the highly deformed zone travels along the gel. For the studied parameters, the peeling front remains straight and stable: no oscillation nor stick-slip motion like those reported by Cortet \& al. \cite{cortet2007imaging} were observed. The peeling dynamics reaches a steady travelling state after a transient that lasts less than a second for the largest velocities. The geometric aspects can be inferred from Fig.~\ref{Fig1}~(b). Interestingly, the typical scales of deformation, $e$ in the normal direction and $\ell$ parallel to the gel layer, depend non-trivially on $f$. In Fig.~\ref{Fig2} we plot the normal displacement $e$ 
as a function of forcing $f$, at different inclination angles $\phi$. The response is far from linear, with a scaling law $e \sim f^{1/2}$, even though typical strains are within the linear range of the gel.

\begin{figure}[t!]
\includegraphics{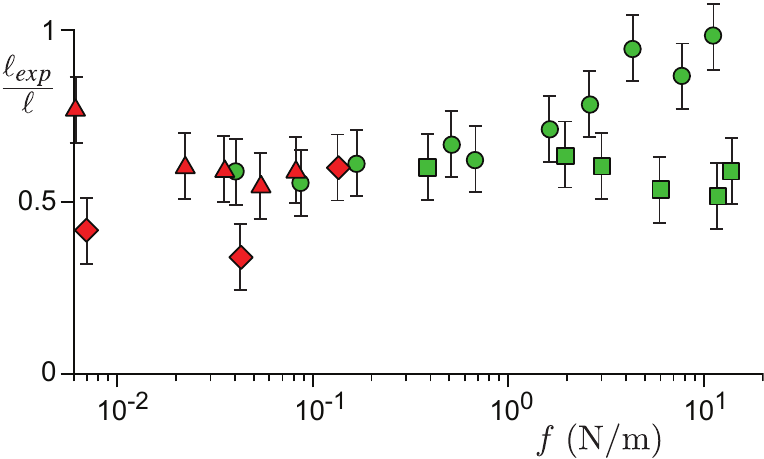}
\caption{Measured lateral decay length $\ell_{exp}$ on the side of the tape divided by the elasto-bending length $\ell$ for different forces. The symbols are the same as on the other figures.}
\label{Fig3}
\end{figure}

This nonlinear response of the normal displacement can be understood in two steps. First, we analyze the part of the elastic gel that is in contact with the tape. The gel layer is soft and thick, and characterized by its (static) shear modulus $G$. A surface deflection of amplitude $e$ and horizontal scale $\ell$ induces a normal stress $\sigma \sim G e/\ell$. By contrast, the elasticity of the thin, stiff tape is characterized by its bending modulus $B$, and gives a typical normal stress $\sigma \sim B e/\ell^4$. The balance of stress thus does not select the normal deflection $e$, but provides access to a lateral length scale $\ell$ given by:
\begin{equation}\label{eq:horizontal}
\ell = \left(\frac{B}{G}\right)^{1/3}.
\end{equation}
This elasto-bending length $\ell$ is the wavelength of wrinkles that appear when compressing a soft foundation that is covered by a hard, thin skin \cite{miquelard2010contact,BD11}. In the present case of a peeling experiment, where the elastic film is under tension, $\ell$ represents the decay length over which the angle of the tape aligns to the gel layer (Fig.~\ref{Fig1}b) ( $\simeq 1.8 \;\rm{mm}$ for the flexible tape and $\simeq 4.3\;\rm{mm}$ for the stiff one). To check this prediction, we have directly measured the length $\ell_{exp}$ over which the tape relaxes to its asymptotic angle. The measurements, reported in Fig.\ref{Fig3} shows that it presents subdominant variations when the force is varied over more than three orders of magnitude. Moreover, it is as expected on the order of the elasto-bending length $\ell$.

\begin{figure}[t!]
\includegraphics{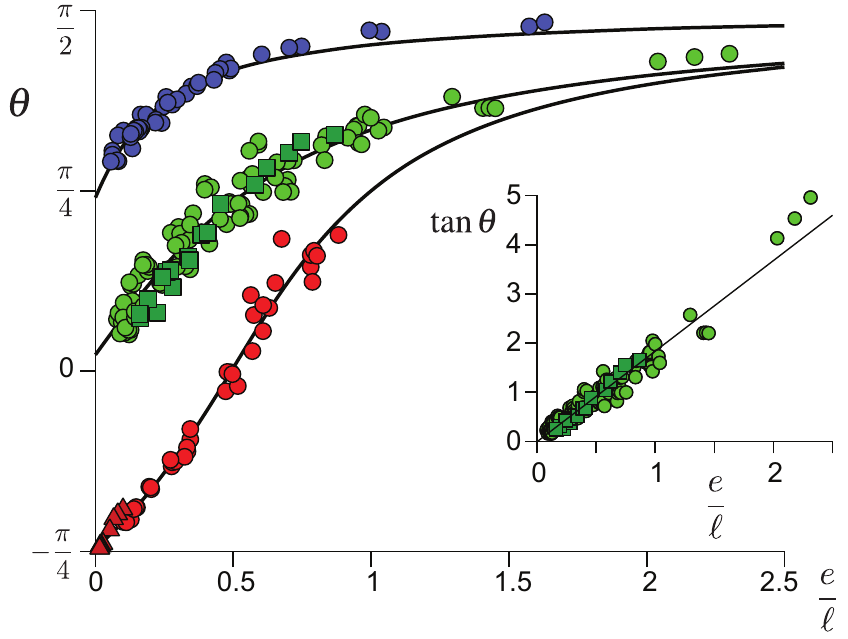}
\caption{Angle $\theta$ at the contact line as a function of the deformation $e$, rescaled by the elastic length $\ell$. The data correspond to different types of tape and configurations. The colors correspond to the global inclination angle $\phi$. The solid lines show the best fit by the form $\tan \theta=\tan \phi+ e/(\alpha \ell)$. For red points $\phi=-45^{\circ}$ and $\alpha=0.51$, for green points $\phi=0^{\circ}$ and $\alpha=0.6$, for blue points $\phi=45^{\circ}$ and $\alpha=0.18$. Inset: slope $\tan \theta$ of the sheet at the contact line for $\phi=0^\circ$. The solid line indicates $\tan \theta \sim e/\ell$.
\label{Fig4}}
\end{figure}

As the length $\ell_{exp}$ is only rigorously defined in the limit of small deformations where an exponential relaxation towards a flat surface nicely fits the data, we have also measured the inclination of the tape at the contact line $\theta$ with respect to the horizontal, for different peeling forces (see Fig.~\ref{Fig1}b for the definition of $\theta$). A very simple view of the deformation would be to approximate the tape as a triangle, with normal extension $e$ and lateral extension on the side of the gel given by $\ell$. This would give a geometric relation $\tan \theta - \tan \phi = (e/\ell) (1 + \tan \theta \tan \phi)$. However, this approximation has the drawback that it does not capture the large deformation asymptotic: when $e\to +\infty$, $\theta \to \pi/2$ since the tape is aligned with gravity already at the location of the contact line. Therefore we propose the fit:
\begin{equation}\label{eq:fit}
\tan \theta - \tan \phi =  \frac{e}{\alpha \ell},
\end{equation}
as it correctly gives the divergence of the normal deformation $e$ when $\theta \to \pi/2$ and is consistent with the triangular model for $\phi=0$. Figure~\ref{Fig4} presents the angle $\theta$ as a function of the deformation $e/\ell$ for the different sets of experiments -- type of tape, configuration (with or without a mass) and global inclination $\phi$. The inset shows the case where the gel is horizontal ($\phi=0$). The data are very well described by the fit (\ref{eq:fit}) for all $\phi$, shown as a solid lines. The prefactor $\alpha$ was found of order unity for all cases (see caption), and shows that $\ell$ as defined by (\ref{eq:horizontal}) indeed sets the lateral scale -- with a weak dependence on the inclination angle $\phi$.  The result obtained for $\phi=0$ is shown in the inset of Fig.~\ref{Fig4} and is accurately described by $\tan \theta \sim e/\ell$, for different $f$ and $B$, confirming that (\ref{eq:horizontal}) correctly predicts the lateral scale $\ell$.

To explain the nonlinear behaviour of the normal displacement $e$, we now turn to the freely hanging part of the sheet that is not in contact with the gel. The shape is that of a classical elastica, forced by the mass at the end of the tape. Here we parametrise the shape of the tape by $\theta(S)$ relating its local angle to the curvilinear coordinate $S$ (Fig.~\ref{Fig2}, inset). Introducing the tangential unit vector $\vec{t} =(\cos \theta,\sin \theta)$, the elastica equation can be integrated to 
\begin{equation}\label{eq:elastica}
\frac{1}{2} B \theta'(S)^2  +  \lambda \vec g \cdot \left[ \vec t(S)-\vec t(R) \right] =0,
\end{equation}
where we used that the end of the tape at $S=R$ is free from torque. At the contact line we can estimate the bending term from the characteristic horizontal and vertical scales: $B \theta'^2 \sim Be^2/\ell^4$. The forcing term in (\ref{eq:elastica}) can be written as $\lambda g(1-\sin \phi)=f$, which becomes exact when the hanging part of the sheet is long. Combined with (\ref{eq:horizontal}), this gives the nonlinear scaling for the normal displacement $e$:
\begin{equation}\label{eq:normal}
\frac{Be^2}{\ell^4} \sim \frac{e^2G}{\ell} \sim f .
\end{equation}
This relation is successfully tested in Fig.~\ref{Fig2}, where we find a collapse over 4 decades, for different inclinations $\phi$, for two different bending moduli $B$, for two different gel elastic moduli G, for different gel thicknesses and for different adhesion conditions. Note that $e/\ell$ provides the order of magnitude of the strain, which shows that most of our experiments are performed in the linear visco-elastic regime.
 
\section{Reversibility, work of adhesion and contact angle}
\begin{figure}[t!]
\includegraphics{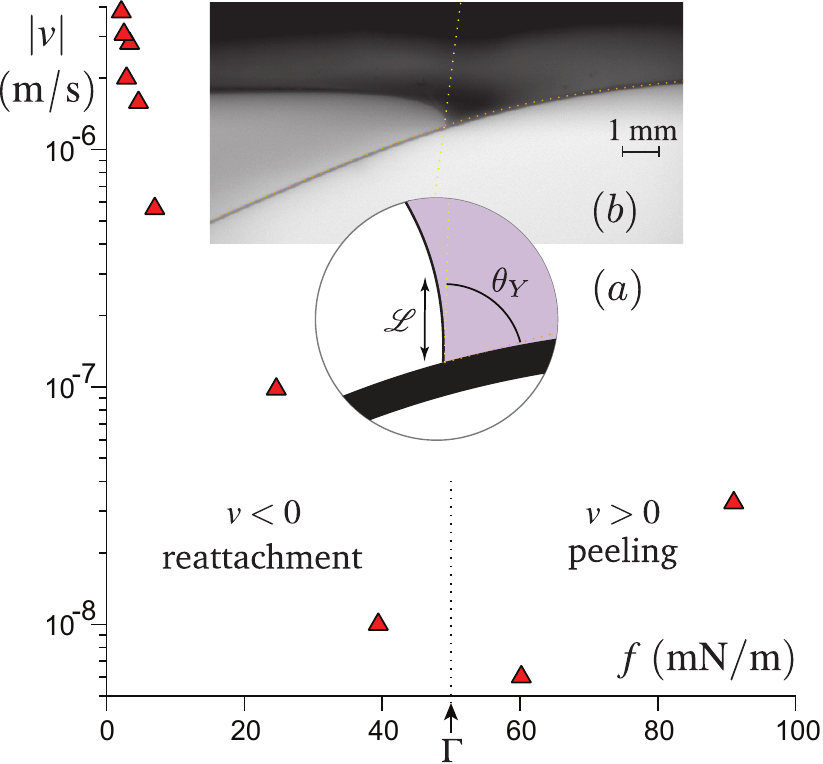}
\caption{(a) Relation between the peeling velocity $|v|$ and the peeling force $f$ on semi-logarithmic axes, for the case where no mass is added to the sheet. Negative velocities, for which the tape reattaches, are on the left, for forces below $\Gamma$. Data correspond to the metallized film and the softer gel. (b) Side view of the gel in the equilibrium condition $f=\Gamma$, showing Young's contact angle $\theta_Y\simeq 65 \pm 5^\circ$.}
\label{Fig5}
\end{figure}

The reversibility of the peeling process is illustrated by a series of experiments without an additional mass at the end of the tape. When the hanging part of the sheet is sufficiently long, peeling can in fact be induced by the weight of the sheet. However, there is a threshold length below which the surface energy due to adhesion is stronger than gravity, so that the sheet spontaneously reattaches to the gel when its end is released. The resulting peeling velocities $v$ are presented in Fig.~\ref{Fig5}a, where positive $v$ correspond to peeling and negative $v$ to reattachment. At the threshold point where $v=0$, the system is at equilibrium: there is an exact balance between the forcing $f$ and the (conservative) work of adhesion $\Gamma$ -- note that this equilibrium configuration is unstable in the sense that if the tape peels off, the peeling force increases, so that the tape continues to peel off. The value for the softer gel and the metallized sheet is $\Gamma= 52 \pm 3 \,{\rm mN.m^{-1}}$  and for the softer gel and the Mylar sheet is $\Gamma= 19 \pm 3 \,{\rm mN.m^{-1}}$. For the ten times more rigid gel and the metallized sheet $\Gamma= 10 \pm 5 \,{\rm mN.m^{-1}}$. Hence, this experiment allows us to accurately determine the non-dissipative contribution to the adhesion process. Note that reversibility is, here, not intended in the thermodynamic sense (there is a visco-elastic dissipation) but in the structural sense: the adhesive does not present irreversible damages after peeling.

The work of adhesion can be directly related to the shape of the gel on the side that is detached from the sheet. Figure~\ref{Fig5}b shows that the free surface of the gel is highly curved before contacting the sheet. Just like in a crack \textit{\`a la} Griffith~\cite{G20} or in the JKR adhesion problem~\cite{JKR71}, the normal deflection follows a square root shape $h \sim x^{1/2}$, where $x$ is the horizontal distance measured from the contact line~\cite{JKR71} (see Appendix \ref{AppendixB}). This square root singularity is the solution of the mixed problem, where a contacting region changes to a stress-free interface. It must not be confused with the logarithmic deformation of a surface submitted to a localized force, which appears here (or in the JKR problem) as the far field solution far away from the contact line \cite{miquelard2010contact}. The multiplicative factor in front of the square root solution is usually selected by matching to the outer problem. Remarkably, however, the gel is clearly seen to make a well-defined contact angle $\theta_Y$ when touching the sheet (Fig.~\ref{Fig5}b). This strongly resembles the wetting of liquids, for which the work of adhesion can be expressed as
\begin{equation}
\label{eq:Young}
\Gamma = \gamma(1+\cos \theta_Y), 
\end{equation}
where $\gamma$ is the surface energy of the liquid-vapor interface and $\theta_Y$ is Young's contact angle. The gel could indeed obey similar wetting laws~\cite{StyleNatComm2013,SalezSM2013,KWS16} below the elastocapillary length 
\begin{equation}
\mathcal L=\frac{\gamma}{G},
\end{equation}
where in this case $\gamma$ is the surface energy of the gel. This length, approximately $30\, \mu{\rm m}$ for the softer gel and ten times lower for the more rigid, indicates the scale where surface energy dominates over bulk elasticity, and below which we expect "wetting" behavior.

Despite subtleties of capillarity of elastic interfaces \cite{Andreotti2016a,AndreottiSM2016,SJHD17,Schulman2018aa}, this wetting interpretation is indeed consistent with our direct measurement of the gel's contact angle. For the softer gel, we measured $\theta_Y= 65\pm 5^\circ$ with the metallized sheet, and a higher contact angle $\theta_Y= 125\pm 5^\circ$ with the less adhesive Mylar sheet. Using the measured $\Gamma$ and the previously reported value for the surface tension of the gel-vapor interface, $\gamma=39 \,{\rm mN.m^{-1}}$ \cite{KarpNcom15}, the aforementioned relation predicts contact angles of respectively $\theta_Y= 70^\circ$ and $\theta_Y= 120^\circ$, consistent with direct optical estimates.

\section{Energy release and dissipation} 
\begin{figure}[t!]
\includegraphics{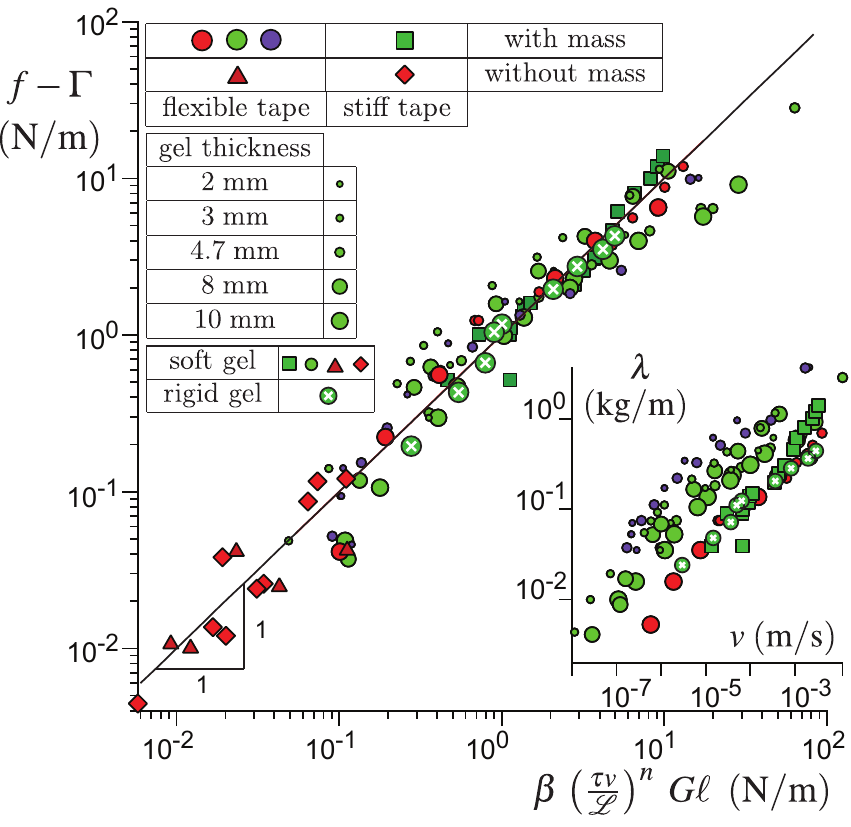}
\caption{Relation between the peeling force $f$ and the peeling velocity $v$, scaled according to (\ref{eq:fdisv2}). The colors correspond to different inclination angles $\phi$, from red to blue ($-45^\circ$, $0^\circ$, $45^\circ$) and the symbols to the type of experiment, as shown in the table in inset. The symbol size codes for gel thickness. The inset represents the data without rescaling. The circled cross symbol corresponds to the $7.5\;\rm{mm}$ thick layer of the most rigid gel and the flexible backing, for $\phi\simeq0^\circ$, with a mass.
}
\label{Fig6}
\end{figure}

We now turn to the most important characteristic of the adhesive, namely the relation between the peeling force and the peeling velocity. The unscaled experimental data is shown in the inset of Fig.~\ref{Fig6}, where we present the mass $\lambda$ \textit{versus} the velocity $v$. For each dataset with given tape and inclination $\phi$, we find a power law with an exponent $0.53 \pm 0.04$ for the softer gel and an exponent $0.37 \pm 0.15$ for the ten times more rigid gel. Below we demonstrate that this directly reflects the exponent $n$ of the rheology, as given by (\ref{eq:gelrheology}). Another observation is that the peeling velocity is independent of the gel thickness (represented by the size of the symbols), from which we deduce that dissipation is localized in the vicinity of the peeling front.
%

To interpret the relation between forcing and peeling velocity, we make use of an energy balance. Since inertial effects are negligible, the viscoelastic dissipation in the bulk of the deformed gel must equal the changes in surface energy (due to the work of adhesion) and the forcing by the weight. We thus need to determine the dissipation inside the bulk of the gel layer due to the rheology (\ref{eq:gelrheology}). A closed form expression for the dissipation can be obtained in the limit of small deformations, assuming that the gel profile (cf. inset Fig.~\ref{Fig2}) is a traveling wave $h(x-vt)$. The resulting balance reads (see Appendix~\ref{sec:dissipation}):
\begin{equation}\label{eq:balance}
f -\Gamma= \int \frac{dq}{2\pi}  \frac{qG''(qv)}{k(q)}  |\hat h(q)|^2.
\end{equation}
The integral on the right hand side represents the dissipation, where $\hat h(q)$ is the Fourier transform of the gel deformation and $k(q)$ is the spatial Green's function relating deformation to the normal stress. The finite thickness $H$ of the gel layer enters in $k(q)$ for $qH\lesssim 1$. In our case deformation is limited to $\ell\lesssim H$ and we may use the half-space approximation $k(q) \sim (2|q|)^{-1}$; in agreement with the experimentally observed independence of thickness. Naturally, $\omega =qv$ sets the characteristic frequency for the dissipation, as can be seen from the argument of $G''$ in (\ref{eq:balance}). Furthermore, the independent calibration (\ref{eq:gelrheology}) allows us to write $G'' \sim  (qv \tau)^n G$.

The final step is to insert the deformation profile $\hat h(q)$ in (\ref{eq:balance}). Importantly, when using the crack-shape $h \sim (\ell x)^{1/2}$, for which $\hat h \sim \ell^{1/2}/|q|^{3/2}$, the dissipation integral in (\ref{eq:balance}) diverges at large $q$ and gives infinite dissipation at small scales (see Appendix \ref{AppendixB}). However, the appearance of a wetting condition at scales below $\sim \mathcal L$ provides a cutoff: dissipation becomes integrable when the interface exhibits a finite angle, with details depending on $\theta_Y$.  Using $q \sim \mathcal L^{-1}$ as the regularisation scale, (\ref{eq:balance}) becomes:
\begin{equation}
f -\Gamma = \beta\,\left(\frac{\tau v}{\mathcal L}\right)^n\;G \ell.
\label{eq:fdisv2}
\end{equation}
Figure~\ref{Fig6} confirms that all data are well-described by (\ref{eq:fdisv2}) with $\beta$ a multiplicative factor encoding the boundary condition effect and higher order large deformation effects. The precise value of $\beta$ primarily depends on the geometry of the gel at the peeling front: it is found to be $\sim 10$ times larger for $\theta_Y\simeq 65^\circ$ than for $\theta_Y\simeq 125^\circ$. This illustrates the importance of the contact angle, also for dynamics, since a larger peeling ridge leads to stronger dissipation. $\beta$ exhibits a subdominant non-monotonic dependence on $\phi$ that can be traced back to a weak variation of the lateral relaxation scale (see the value of $\alpha$ in Fig.~\ref{Fig4}).
\begin{table}[h!]
\centering
\begin{tabular}{|c|c|cc}
\hline
$ \beta$  &   $\phi \simeq 0^\circ$& \multicolumn{1}{c|}{$\phi \simeq -45^\circ$} & \multicolumn{1}{c|}{$\phi \simeq 45^\circ$} \\ \hline
$ \theta_Y = 65^\circ \pm 5^\circ$ & 3.3 & \multicolumn{1}{c|}{1.2-1.4} & \multicolumn{1}{c|}{2.2} \\ \hline
$ \theta_Y = 125^\circ \pm 5^\circ$ & 0.43  & \multicolumn{1}{c|}{0.07} &                       \\ \cline{1-3}
$ \theta_Y = 160^\circ \pm 15^\circ $ &0.25  &                       &                       \\ \cline{1-2}
\end{tabular}
\caption{Table of the parameter $\beta$ defined by the force balance~\eqref{eq:fdisv2} according to the various experimental configurations: type of tape and inclination $\phi$.}
\end{table}

\section{Discussion}

While first theories of peeling of pressure sensitive adhesives have described purely static equilibrium situations~\cite{kaelble1959,kaelble1960}, it was soon realised that the peel force was velocity dependent and associated to the  rheology of the adhesive~\cite{kaelble1964,Gent69,kendall1971adhesion,kendall1975thin}. The problem of growing cracks in bulk viscoelastic media is closely related, and \citet{schapery1975theory} predicted a power law dependence of the growth rate of the crack on the applied gross strain, where the exponent is given by the exponent of the creep compliance function. Also closely related is the geometry of a rigid cylinder rolling on a viscoaleastic material, for which a strain energy release rate $\sim v^{0.55}$ has been found~\cite{barquins1988jadhesion,barquins1998intj}. \citet{maugis1978} showed that peeling of urethane strips off a glass surface yields a power law relation between peeling speed and applied force as well. They attributed this behavior to the rheology of the adhesive since the empiric dissipation relation obeyed the same time-temperature superposition behavior as the loss modulus. They also noticed that, if viscoelastic losses are localized at the crack tip, dissipation was supposed to be geometry independent.

However, to the best of our knowledge, a quantitative theory capable of predicting the peel force for a highly localized dissipation was still missing in literature. Historically, this is most likely due to the fact that typical properties of engineered adhesive tapes lead to characteristic length scales which do not admit such strong localisation: Such tapes typically comprise a thin layer of a soft and highly adherent material, so that the stress localisation $\sim\Gamma/G\sim \mathcal{O}(1~\text{mm})$ is spread much wider than the typical layer thickness $\sim\mathcal{O}(10~\upmu\text{m})$~\cite{VC15,CC16}. Instead, a cohesive zone forms, composed of fibrils and cavitation bubbles, and the strain energy release rate becomes geometry dependent and typically increases with layer thickness~\cite{kaelble1992}. Those cases clearly require a different type of model, taking the nonlinear rheological properties of the adhesive into account~\cite{vilmin2009simple,nase2008,CC16}.

The growing interest in reversible adhesives that can be peeled without bulk cavitation or plastic deformations, as inspired e.g. by biology~\cite{ghatak2004peeling,LW14}, suggests revisiting the limit of weakly adherent materials and peeling geometries with localised stress and dissipation.
However, the crack singularity of adhesion leads in this case not just to a concentration of the stress to the peeling front, but also to a diverging dissipation in the continuum description. This prevents quantitative predictions unless a physical regularisation mechanism is identified. While the power laws that we find in our experiments are similar to those reported previously~\cite{maugis1978,barquins1998intj}, we could disentangle the various contributions to the peel force by peeling an inextensible flexible tape with different surface energies off a thick elastic layer~\cite{ghatak2005measuring}. Most importantly we have identified solid surface tension and the corresponding ``wetting type'' boundary condition as the leading order regularisation mechanism for dissipation in reversible peeling. 

In conclusion, we have shown that reversible adhesives can indeed obey simple scaling laws for deformation and dissipation, whose origin can be traced back to linear viscoelasticity and, importantly, solid capillarity. Analysing the near-crack-tip geometry, we have shown that the regularisation of elastic singularities by the wetting condition allows one to get  quantitative estimates of the viscoelastic dissipation. The theoretical framework proposed here bridges the gap between adhesive peeling and moving contact lines of fluids, opening the possibility of fully quantitative theories. It opens the promising perspective of designing adhesives by coherently tuning their visco-elastic properties, their surface functionalisation, and their meso- and macroscopic architecture.

\section{Appendix: derivation of the model equations}
In this appendix section, we derive the model used along the article as a framework of interpretation of experimental results. We provide at each step the general, rigorous equations, and their expansion at the lowest order in deformation and contact line velocity.

\subsection{Peeling force and boundary condition at the contact line}\label{app:weight}
\label{AppendixA}
Here we will derive the expression of the peeling force $f$, for the experiments with or without and additional mass attached at the end of the tape. The length $R$ of the tape is arbitrary. The driving force of the peeling motion is then due to the weight of the tape, the additional mass, and the bending moment of the tape. To derive the governing equations for the shape of the free part of the tape, we use curvilinear coordinate $S$ and the corresponding tangent vector $\vec{t}(S) = \cos \theta(S) \vec e_x+\sin \theta(S) \vec e_y$ along the contour of the tape; $\theta(S)$ denotes the angle between the tangent vector and the horizontal. We define that the contact line position corresponds to the curvilinear coordinate $S=0$, and the end of the tape is located at $S=R$. The torque balance on the piece of tape going from $S$ to $R$ reads:
 \begin{equation}
B\theta'(S) \vec e_z+ \rho_S  \vec g \times \int_S^R (R-S') \vec t(S') dS'+ \lambda \vec g \times \int_S^R \vec t(S') dS' =0  \label{eqtorque}
 \end{equation}
The first term, which involves the bending modulus $B$ of the tape, is the elastic torque. The second term is the gravity torque due to the weight of the tape itself. $\rho_s$ is the mass density of the tape per unit area and $\vec g=g( \sin \phi \vec e_x + \cos \phi \vec e_y)$ is the gravitational vector. The last term is the gravity torque due to the additional mass at the end of the tape, $\lambda$, per unit width of the tape.

\begin{figure}[t!]
\includegraphics{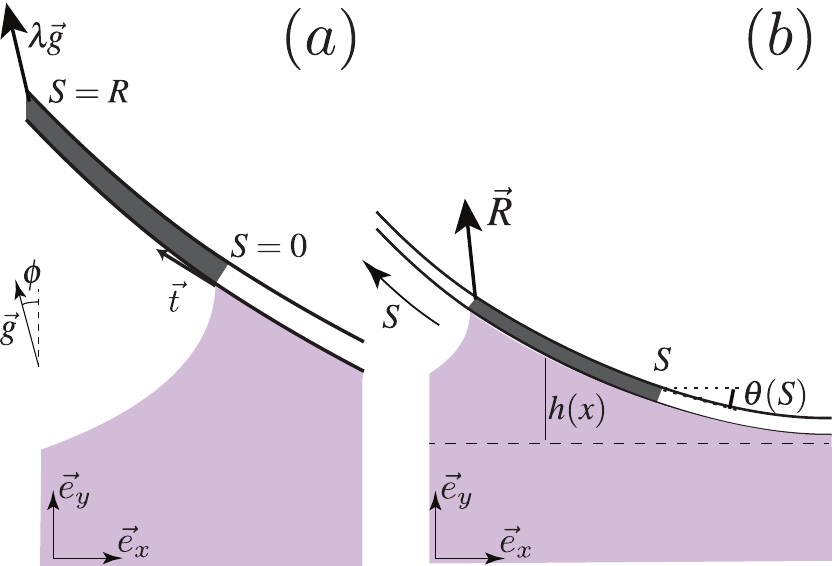}
\caption{Theoretical schematics. (a) For the derivation of the peeling force $f$ we only consider the hanging tape (shaded on the scheme) for which the contour coordinate $S>0$. (b) For the derivation of gel-tape equilibrium shape we consider the attached part of the tape and the elastocapillary mechanics of the gel. The shaded part of the attached tape represents the system considered for the derivation from $0$ to $S<0$.}
\label{Fig7}
\end{figure}

We define the peeling force $f$ as the energy released by gravity and bending; here we ignore -- for now -- the surface tensions of tape and gel. We therefore write the free energy $\mathcal{F}_{d}$ of the free part of the tape and calculate its variations with respect to shape and contact line motion. The bending energy and gravitational energy of the detached part of the sheet are expressed as:
\begin{eqnarray}
\mathcal{F}_{d}\!&=&\! \int_{0}^{R} dS \frac{1}{2} B  \theta'(S)^2\nonumber\\
&+&\frac12\rho_S R^2 g \sin \phi - \int_{0}^{R} dS \rho_S \vec{g} \cdot \int_0^S \vec t (S')dS'\nonumber\\
&+&\lambda R g \sin \phi- \lambda \vec g \cdot \int_{0}^{R} dS\,\vec t(S)
\label{eqFd}
\end{eqnarray}
The first line of the right hand side is the bending energy, the second line describes the gravitational energy due to the weight of the tape itself and the third line is the gravitational energy of the additional mass at the end of the tape. Note that both gravitational terms result from two contributions each: not only the total hanging contour length $R$ of the tape changes (second term), but also the contact line position moves simultaneously (first term). Here we ignore that the gel deformation also slightly depends on $R$, which causes an additional motion of the contact line relative to the laboratory frame. Now, the total variation of the free energy with respect to a change of length $R$ gives the peel force $f$:
\begin{eqnarray}
	f \!&=&\! - \frac{\partial \mathcal{F}_{d}}{\partial R}\\
	    &=&\! - g(\rho_S R + \lambda)\sin\phi 
	    			  + \lambda \vec g \cdot \vec t(R)
	    			  + \rho_S \vec{g} \cdot \int_{0}^{R}\!dS\;\vec t (S)
\nonumber
\end{eqnarray}
To obtain $f$ from the experiment, all terms on the right are evaluated directly from the images. We have checked experimentally that the variation of the angle $\theta$ at the contact line with respect to $R$ leads to a negligible contribution to $f$.
\begin{figure}[t!]
\includegraphics{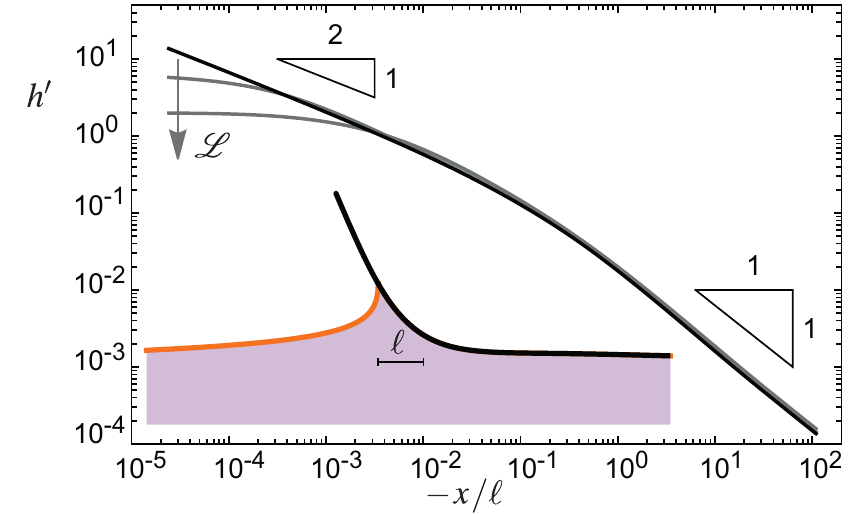}
\caption{Predictions of the linear visco-elastic model (equations
~(\ref{eq:normalstress})-(\ref{eq:hilbert})) at vanishing velocity $v$. Typical equilibrium profiles of the slope $h'$ on the free side of the contact line as a function of the coordinate $x$ rescaled by $\ell$ for three different contact angles. At small $x \lesssim \mathcal L$, $h'$ saturates to the value selected by the balance of surface tensions. For $\mathcal L \lesssim x \lesssim \ell$, an intermediate asymptote $h'\sim x^{-1/2}$ emerges. Therefore, $h\sim x^{1/2}$, reminiscent of the crack opening problem. Finally, at large $x\gtrsim \ell$, one recovers an outer asymptote $h'\sim x^{-1}$ or equivalently $h\sim\log x$, which is the elastic response to a localised line load. Inset: complete solution $h(x)$ showing the relaxation over a scale $\ell$ on the tape side and the intermediate asymptotics featuring a square root behaviour on the free side.}
\label{Fig8}
\end{figure}

\subsection{Equations coupling the tape to the gel mechanical response}
\label{AppendixB}
In order to derive the equations governing the shape of the gel surface, we proceed to a double expansion: we consider the rheology as being linear, which is well satisfied by the gel used here, even under stretching; furthermore, we linearize the equations at small deformations of the gel surface, in order to be able to use the Green's function formalism. Although this linearity requirement is badly obeyed in many of our experiments, we argue here that the laws found using this approximation may still be valid, within a multiplicative constant that cannot be calculated quantitatively. The scaling behaviour of our experimental results show that this is indeed the case. We furthermore neglect gravity in the description of the gel.

We first introduce the elastic stress $\sigma$ at the surface of the gel. On the free surface side of the gel, 
the shear stress vanishes, and the normal stress is balanced by the solid Laplace pressure $\Upsilon_{S} h''$. The solid surface tension $\Upsilon_{S}$ is considered constant, for simplicity ($\Upsilon_{S} = \gamma = \mathcal{L} G$). On the tape side, the strain $\epsilon$ along the surface vanishes on the tape side. This implies that, at linear order, the shear stress vanishes, but not the normal stress. The latter is set by the bending of the tape, $B h''''$.
Thus,
\begin{equation}
	\label{eq:normalstress}
	\sigma = \left\{\begin{array}{ll}
		\Upsilon_{S} h'' & \text{for}\quad x<0\\
		-B h'''' & \text{for}\quad x>0
	 \end{array}\right.
\end{equation}
There are four boundary conditions at the contact line: the selection of Young's angle (Eq.~\ref{eq:Young}), the selection of the tape angle $\theta(0)$, and the second and third derivatives inherited from the free part of the tape.

Next we relate the normal traction $\sigma$ to the normal displacement and to the viscoelastic rheology  $\mu(\omega)=G'(\omega)+iG''(\omega)$. When taking the Fourier transform in both space ($x \rightarrow q$) and time ($t \rightarrow \omega$), the kernel relating the deformation to the normal stress can be written as the ratio $\mu(\omega)/k(q)$ \cite{LAL96,KarpNcom15}, where $k(q)$ is the spatial Green's function corresponding to the linear response. Here we implicitly made use that for an incompressible thick layer, the normal stress is decoupled from the tangential displacement\cite{Johnson}. We consider travelling wave solutions which depend on the single variable $x-vt$. The Fourier transforms of the elastic stress $\hat \sigma$ and the gel profile $\hat h$ are then related by:
%
%
\begin{equation}
\hat \sigma(q) = \frac{\mu(qv)}{ k(q)} \hat h(q)
\label{Eqsigdyn}
\end{equation}
The effect of velocity and surface tension are therefore only to affect the space Green's function. At small velocity $v$, the rheology can be expanded at the lowest order, leading to the static shape relation. In this case and for a thick layer ($\ell\ll$ gel thickness), the profile slope and the stress are related to each other by a Hilbert transform:
\begin{equation}
	\label{eq:hilbert}
	\sigma(x) = -\frac{2G}{\pi}\int_{-\infty}^{\infty}\frac{h'(x')}{x'-x}dx'
\end{equation}
Considering previous work for liquid droplets on the same gel\cite{KarpNcom15}, the range of velocities for which this is valid extends up to \hbox{$2\cdot 10^{-3}\;{\rm m/s}$.
 %
 %
The above set of equations} is complemented by the asymptotic condition $h'\sim -x^{-1}$ for $|x|\gg \ell$, which corresponds to load that is localized on $|x|\ll\ell$.

Figure \ref{Fig8} shows the typical surface profiles obtained  by numerical resolution of the visco-elastic model in the limit of small velocities. Far from the contact line, the shape becomes logarithmic, as expected from the response to a localized force \cite{miquelard2010contact}. At intermediate distances to the contact line, the normal deflection follows the square root shape $h \sim \left(\ell x\right)^{1/2}$, as is expected in such a mixed problem where a compliant contact zone connects to a stress-free surface~\cite{G20,JKR71}. The lengthscale scale $\ell$ appearing in front of the square root singularity is selected by matching to the outer solution~\cite{C5SM03079J}. However, due to solid capillarity, the square root singularity is regularized close to the contact line, leading to a well-defined Young's contact angle. The square root shape is retained as an intermediate asymptote.

\subsection{The dissipation integral\label{sec:dissipation}}
Here we compute the rate energy dissipation $P$ inside the gel during peeling, which is used to derive equation (\ref{eq:balance}) in the main manuscript. We define the displacement field inside the gel $u_i$ and the stress tensor $\sigma_{ij}$. With this, we compute the energy dissipation per unit time (per unit width of the gel) from the usual integral \cite{LAL96}
\begin{eqnarray}\label{eq:diss}
P &=& \int d^2x \,  \sigma_{ij} \frac{\partial \dot u_i}{\partial x_j}  \nonumber \\
&=& \int d^2x \,  \left[ \frac{\partial }{\partial x_j} \left( \sigma_{ij} \dot u_i\right) - \dot u_j \frac{\partial \sigma_{ij}}{\partial x_j} \right] \nonumber \\
&=& \oint ds \,    \sigma_{ij} n_j \dot u_i.
\end{eqnarray}
These contain standard manipulations, where we used mechanical equilibrium $\partial \sigma_{ij}/\partial x_j=0$ and brought the area integral to the boundary with normal vector $n_j$. The boundary integral represents the work done by normal and tangential tractions. Since the bottom of the gel is fixed to a rigid support ($\dot u_i=0$) the only contribution comes from the free surface. In addition, only the normal traction performs work: the inextensibility of the sheet imposes vanishing tangential displacement, while the tangential stress vanishes on the side that is peeled from the sheet. In the limit of small deformation, the normal displacement can be identified with the gel's profile $h(x,t)$, so that (\ref{eq:diss}) reduces to 
\begin{eqnarray}\label{eq:eqdiss3}
P &=& \int_{-\infty}^\infty dx\, \sigma(x,t) \dot h(x,t),
\end{eqnarray}
where we denote the normal stress $\sigma=\sigma_{ij} n_i n_j$.

Inserting in (\ref{eq:eqdiss3}), the dissipation becomes:
\begin{eqnarray}
P &=& \int dx \left\{ 
\int \frac{dq}{2\pi}  \frac{\mu(qv)}{k(q)}h_c(q) e^{iq(x-vt)} 
\right\}  \nonumber \\
&& \times
\left\{ 
\int \frac{dq'}{2\pi} (iq'v) h_c(q') e^{iq'(x-vt)} 
\right\}
\end{eqnarray}
We can introduce the variable $\tilde x = x-vt$ and perform the $x$ integral using
\begin{equation}
\int d\tilde x \, e^{i(q+q') \tilde x} = 2\pi \delta(q+q').
\end{equation}
This gives a dissipation
\begin{eqnarray}
P &=&\int \frac{dq}{2\pi}  \frac{\mu(qv)}{k(q)}h_c(q)  
\int \frac{dq'}{2\pi} (iq'v) h_c(q')\, 2\pi \delta(q+q') \\
&=&- v \int \frac{dq}{2\pi}  \frac{\mu(qv)}{k(q)}h_c(q)   (iq) h_c(-q) \\
&=&- v \int \frac{dq}{2\pi}  \frac{\mu(qv)}{k(q)} (iq) |h_c(q)|^2.
\end{eqnarray}
This expression is real, owing to the symmetry properties $k(q)=k(-q)$ and $\mu(-\omega)=\mu(\omega)^*$, which implies $G'(\omega)=G'(-\omega)$ and $G''(\omega)=-G''(-\omega)$. Hence, as expected, only $G''$ contributes to the dissipation. Finally, we can define the dissipative force $P=f_d v$. This gives the expression used in equation (\ref{eq:balance})
\begin{equation}
f_d =  \int_{-\infty}^\infty \frac{dq}{2\pi}  \frac{q G''(qv)}{k(q)}|h_c(q)|^2.
\end{equation}

{\bf Acknowledgments.} 
B.A. is supported by Institut Universitaire de France. This work was funded by the ANR grants Smart. JHS acknowledges financial support from ERC (the European Research Council) Consolidator Grant No. 616918. SK acknowledges financial support from the Max Planck -- University of Twente Center `Complex Fluid Dynamics -- Fluid Dynamics of Complexity'.


\providecommand*{\mcitethebibliography}{\thebibliography}
\csname @ifundefined\endcsname{endmcitethebibliography}
{\let\endmcitethebibliography\endthebibliography}{}

\end{document}